\documentclass[twocolumn]{autart}    

\usepackage{graphicx}          

\graphicspath{{img/}}

\usepackage{amssymb}
\usepackage{amsmath}
\usepackage{natbib}

\newtheorem{theorem}{Theorem}

\newtheorem{proposition}{Proposition}
\newtheorem{definition}{Definition}
\newtheorem{remark}{Remark}

\date{8 September 2016}

\begin{document}

\begin{frontmatter}

\title{Contraction analysis of switched systems via regularization\thanksref{footnoteinfo}} 

\thanks[footnoteinfo]{This paper was not presented at any IFAC 
meeting. Corresponding author Mario di Bernardo. Tel. +39-081-7683909. Fax +39-081-7683186.}

\author[Napoli]{Davide Fiore}\ead{davide.fiore@unina.it},             
\author[Bristol]{S. John Hogan}\ead{s.j.hogan@bristol.ac.uk},  
\author[Napoli,Bristol]{Mario di Bernardo}\ead{mario.dibernardo@unina.it}

\address[Napoli]{Department of Electrical Engineering and Information Technology, University of Naples Federico II, Via Claudio 21, 80125 Naples, Italy}  
\address[Bristol]{Department of Engineering Mathematics, University of Bristol, BS8 1TR Bristol, U.K.}

\begin{keyword}                         
contraction theory \sep incremental stability \sep switching controllers \sep regularization   
\end{keyword}

\begin{abstract}                          
We study incremental stability and convergence of switched (bimodal) Filippov systems via contraction analysis. In particular,  by using results on regularization of switched dynamical systems, we derive sufficient conditions for  convergence of any two trajectories of the Filippov system between each other within some region of interest. We then apply these conditions to the study of different classes of Filippov systems including piecewise smooth (PWS) systems, piecewise affine (PWA) systems and relay feedback systems. We show that contrary to previous approaches, our conditions allow the system to be studied in metrics other than the Euclidean norm. The theoretical results are illustrated by numerical simulations on a set of representative examples that confirm their effectiveness and ease of application.
\end{abstract}

\end{frontmatter}

\section{Introduction}
\label{sec:intro}
Incremental stability has been established as a powerful tool to prove convergence in nonlinear dynamical systems \citep{angeli2002lyapunov}. It characterizes asymptotic convergence of trajectories with respect to one another rather than towards some attractor known a priori. Several approaches to derive sufficient conditions for a system to be incrementally stable have been presented in the literature \citep{angeli2002lyapunov,russo2010global,lohmiller1998contraction,forni2014differential,pavlov2006uniform}.

A particularly interesting and effective approach to obtain sufficient conditions for incremental stability of nonlinear systems comes from contraction theory \citep{lohmiller1998contraction,jouffroy2005some,aminzare2014contraction}. A nonlinear system is said to be {\it contracting} if initial conditions or temporary state perturbations are forgotten exponentially fast, implying convergence of system trajectories towards each other and consequently towards a steady-state solution which is determined only by the input (the \emph{entrainment} property, e.g. \citet{russo2010global}). A vector field can be shown to be contracting over a given $K$-reachable set by checking the uniform negativity of some matrix measure $\mu$ of its Jacobian matrix in that set \citep{russo2010global}. Classical contraction analysis requires the system vector field to be continuously differentiable.

In this paper, we consider an important class of non-differentiable vector fields known as \emph{piecewise smooth} (PWS) \emph{systems} \citep{filippov1988differential}. A PWS system consists of a finite set of ordinary differential equations
\begin{equation}
\label{eq:pws}
\dot{x}=f_i(x), \quad x\in \mathcal{S}_i\subset \mathbb{R}^n
\end{equation}
where the smooth vector fields $f_i$, defined on disjoint open regions $\mathcal{S}_i$, are smoothly extendable to the closure of $\mathcal{S}_i$. The regions $\mathcal{S}_i$ are separated by a set $\Sigma$ of codimension one called the \textit{switching manifold}, which consists of finitely many smooth manifolds intersecting transversely.  The union of $\Sigma$ and all $\mathcal{S}_i$ covers the whole state space $U \subseteq \mathbb{R}^n$.

Piecewise smooth systems are of great significance in applications, ranging from problems in mechanics (friction, impact) and biology (genetic regulatory networks) to variable structure systems in control engineering (sliding mode control \citep{utkin2013sliding}) -- for an overview see the monograph by \citep{bernardo2008piecewise}.

The theoretical study of PWS systems is important. Firstly, the classical notion of solution is challenged in at least two distinct ways. When the normal components of the vector fields either side of $\Sigma$ are in the \textit{same} direction, the gradient of a trajectory is discontinuous, leading to Carath\'eodory solutions \citep{filippov1988differential}. In this case, the dynamics is described as \textit{crossing} or \textit{sewing}. But when the normal components of the vector fields on either side of $\Sigma$ are in the \textit{opposite} direction, a vector field on $\Sigma$ needs to be defined. The precise choice is not unique and depends on the nature of the problem under consideration. One possibility is the use of differential inclusions. Another choice is to adopt the Filippov convention \citep{filippov1988differential}, where a \textit{sliding} vector field $f^s$ is defined on $\Sigma$. In this case, the dynamics is described as \textit{sliding}.

Some results have been presented in the literature to extend contraction analysis to non-differentiable vector fields. An extension to piecewise smooth continuous (PWSC) systems was outlined in \citep{lohmiller2000nonlinear} and formalized in \citep{di2014contraction}. Contracting hybrid systems were analysed in \citep{lohmiller2000nonlinear} while the stability analysis of hybrid limit cycles using contraction was presented in \citep{tang2014transverse}. An extension of contraction theory, related to the concept of \emph{weak} contraction \citep{sontag2014three}, to characterize incremental stability of sliding mode solutions of planar Filippov systems was first presented in \citep{di2013incremental} and later extended to $n$-dimensional Filippov systems in \citep{di2014incremental}. Finally, incremental stability properties of piecewise affine (PWA) systems were discussed in \citep{pavlov2007convergence} in terms of \emph{convergence}, a stability property related to contraction theory \citep{pavlov2004convergent}.

In this paper, we take a different approach to the study of contraction in $n$-dimensional Filippov systems than the one taken in \citep{di2013incremental,di2014incremental}. In those papers, the sliding vector field $f^s$ was assumed to be defined everywhere and then the contraction properties of its projection onto the switching manifold was considered (together with a suitable change of coordinates). In the current paper, we adopt a new generic approach which directly uses the vector fields $f_i$ and does not need the explicit computation of the sliding vector field $f^s$. Our method has a simple geometric meaning and, unlike other methods, can also be applied to nonlinear PWS systems. 

Instead of directly analysing the Filippov system, we first consider a {\it regularized} version; one where the switching manifold $\Sigma$ has been replaced by a boundary layer of width $2\varepsilon$. We choose the regularization method of Sotomayor and Teixeira \citep{sotomayor1996regularization}. We then apply standard contraction theory results to this new system, before taking the limit $\varepsilon \rightarrow 0$ in order to recover results that are valid for our Filippov system.
%
%
\section{Mathematical preliminaries and background}
\label{sec:background}
\subsection{Matrix measures}
\label{app:matrix_measure}
Given a real matrix $ A \in\mathbb{R}^{n \times n}$ and a norm $|\cdot |$ with its induced matrix norm $\lVert\cdot\rVert$, the associated \emph{matrix measure} (also called logarithmic norm \citep{dahlquist1958stability,lozinskii1958error,strom1975logarithmic}) is the function $\mu:\mathbb{R}^{n \times n}\rightarrow \mathbb{R}$ defined as 
\begin{equation*}
\mu( A ):=\lim_{h \rightarrow 0^+}\frac{\lVert I+h A  \rVert-1}{h}
\end{equation*}
where $I$ denotes the identity matrix.
%
The following matrix measures associated to the $p-$norm for $p=1,2,\infty$ are often used
\begin{align*}
\mu_{1}( A )=\max_{j}\left[a_{jj}+\sum_{i\ne j}|a_{ij}|\right]\\
\mu_{2}( A )=\lambda_{max}\left(\frac{ A + A^T }{2}\right)\\
\mu_{\infty}( A )=\max_{i}\left[a_{ii}+\sum_{j\ne i}|a_{ij}|\right]
\end{align*}
The matrix measure $\mu$ has the following useful properties \citep{vidyasagar2002nonlinear,desoer1972measure}:
\begin{enumerate}
\item
\label{app:mu_prop2}
$\mu(I)=1$, $\mu(-I)=-1$.
\item
\label{app:mu_prop3}
If $A=\varnothing$, where $\varnothing$ denotes a matrix with all entries equal to zero, then $\mu(A)=0$.
\item
\label{app:mu_prop4}
$-\lVert A\rVert \leq -\mu(-A)\leq Re\,\lambda_i(A) \leq \mu(A)\leq \lVert A\rVert $ for all $i=1,\, 2,\dots, n$, where $Re\,\lambda_i(A)$ denotes the real part of the eigenvalue $\lambda_i(A)$ of $A$.
\item
\label{app:mu_prop5}
$\mu(c\,A)=c\,\mu(A)$ for all $c\geq 0$ (positive homogeneity).
\item
\label{app:mu_prop6}
$\mu(A+B)\leq \mu(A)+\mu(B)$ (subadditivity).
\item
\label{app:mu_prop7}
Given a constant nonsingular matrix $Q$, the matrix measure $\mu_{Q,i}$ induced by the weighted vector norm $\vert  x  \vert_{Q,i}=\vert Qx \vert_i $ is equal to $\mu_i( QAQ^{-1})$. 
\end{enumerate}

The following theorem can be proved \citep{vidyasagar1978matrix,aminzare2014contraction}.
\begin{theorem}
\label{app:thm:weighted_mu}
There exists a positive definite matrix $P$ such that $PA+A^TP<0$ if and only if $\mu_{Q,2}(A)<0$, with $Q=P^{1/2}$.
\end{theorem}
We now present results on the properties of matrix measures of rank-1 matrices, since we will need these in the sequel. We believe that Lemma \ref{lemma:measure_rank1} is an original result. For any two vectors $x,y\in\mathbb{R}^n$, $x,y\neq 0$, the matrix $A=xy^T$ has always rank equal to 1. This can be easily proved observing that $xy^T=[y_1x\; y_2x\;\dots \; y_nx]$.
\begin{proposition}
For any two vectors $x,y\in\mathbb{R}^n$, $x,y\neq 0$ and for any norm we have that $\mu(xy^T)\geq 0$.
\end{proposition}
\begin{pf*}{Proof.}
The proof follows from property \ref{app:mu_prop4} of matrix measures as listed above, that is, for any matrix and any norm $\mu(A)\geq Re\,\lambda_i(A)$, for all $i$, where $Re\,\lambda_i(A)$ denotes the real part of the eigenvalues $\lambda_i(A)$ of $A$. Therefore, since a rank-1 matrix has $n-1$ zero eigenvalues its measure cannot be less than zero.
\end{pf*}
The following important result holds for the measure of rank-1 matrices induced by Euclidean norms.
\begin{lem}
\label{lemma:measure_rank1}
Consider the Euclidean norm $|\cdot|_{Q,2}$, with $Q=P^{1/2}$ and $P=P^T>0$. For any two vectors $x,y\in\mathbb{R}^n$, $x,y\neq 0$, the following result holds
\begin{equation*}
\mu_{Q,2}(xy^T)=0 \quad \mbox{if and only if} \quad Px=-a\,y, \; a>0,
\end{equation*}
otherwise $\mu_{Q,2}(xy^T)>0$.
\end{lem}
\begin{pf*}{Proof.}
Firstly we prove that $\mu_2(xy^T)=0$ if and only if $x$ and $y$ are antiparallel, i.e. $x=-a\,y$ for some $a>0$. Indeed, from the definition of Euclidean matrix measure, $\mu_2(xy^T)$ is equal to the maximum eigenvalue of the symmetric part ${A_s \equiv (A+A^T)/2}$ of the matrix $A=xy^T$.  The characteristic polynomial $p_\lambda(A_s)$ of $A_s$ is \citep[Fact 4.9.16]{bernstein2009matrix}
\begin{equation*}
\begin{split}
p_\lambda(A_s)=& \lambda^{n-2}\left\{\lambda^2-x^Ty\lambda-\frac{1}{4}\left[x^Txy^Ty-x^Tyy^Tx\right]  \right\}\\
=& \lambda^{n-2}\left\{\lambda^2-x^Ty\lambda-\frac{1}{4}\left[|x|^2_2 |y|^2_2-(x^Ty)^2\right]  \right\}.
\end{split}
\end{equation*}
This polynomial has always $n-2$ zero roots and (in general) two further real roots. It can be easily seen from Descartes' rule that their signs must be opposite. Therefore, the only possibility for them to be nonpositive is that one must be zero while the other is negative. Using again Descartes' rule, this obviously happens if and only if $x$ and $y$ are antiparallel.

Now, assume that $\mu_{Q,2}(xy^T)=0$ then, using property \ref{app:mu_prop7} of matrix measures, we have
$
\mu_{Q,2}(xy^T)=\mu_{2}\left(Qxy^TQ^{-1}\right)=\mu_2\left(Qx(Q^{-1}y)^T  \right)=0,
$
and, from the result proved above, $Qx$ and $Q^{-1}y$ must be antiparallel, i.e. $Qx=-a\,Q^{-1}y$ for some $a>0$, or equivalently $Px=-a\,y$.

To prove sufficiency, suppose that $Px=-a\,y$, $a>0$, then $Qx=-a\,Q^{-1}y$ and therefore, using again the result above, we have
$
\mu_{Q,2}(xy^T)=\mu_2(Qxy^TQ^{-1})=a^{-1}\mu_2(-Qx(Qx)^T)=0.
$
\end{pf*}
Note that when $x$ or $y$ (or both) are equal to 0 then by property \ref{app:mu_prop3} of matrix measures $\mu(xy^T)=0$.
\subsection{Incremental stability and contraction theory}
\label{sec:incr_stab}
Before starting our analysis for PWS systems, we present some key results on the contraction properties of smooth systems. Let $U\subseteq\mathbb{R}^n$ be an open set. Consider the system of ordinary differential equations
\begin{equation}
\label{eq:dynamical_sys}
\dot{x}=f(t,x)
\end{equation}
where $f$ is a continuously differentiable vector field defined for $t\in[0,\infty)$ and $x\in U$, that is $f\in C^1(\mathbb{R}^+\times U,\mathbb{R}^n)$.
We denote by $\psi(t,t_0,x_0)$ the value of the solution $x(t)$ at time $t$ of \eqref{eq:dynamical_sys} with initial value $x(t_0)=x_0$.
We say that a set $\mathcal{C}\subseteq \mathbb{R}^n$ is \emph{forward invariant} for system \eqref{eq:dynamical_sys}, if $x_0 \in \mathcal{C}$ implies $\psi(t,t_0,x_0) \in \mathcal{C}$ for all $t\ge t_0$.
\begin{definition}
\label{def:incr_stability}
Let $\mathcal{C}\subseteq\mathbb{R}^n$ be a forward invariant set and $|\cdot|$ some norm in $\mathcal{C}$. System \eqref{eq:dynamical_sys} is said to be \emph{incrementally exponentially stable} in $\mathcal{C}$ if for any two solutions $x(t)=\psi(t,t_0,x_0)$ and $y(t)=\psi(t,t_0,y_0)$ there exist constants $K\geq 1$ and $c>0$ such that $\forall t \geq t_0,\;\forall x_0,y_0 \in \mathcal{C}$
\begin{equation}
\label{eq:ies}
\lvert x(t)-y(t)\rvert \leq  K\, e^{-c(t-t_0)}\,\lvert x_0-y_0\rvert.
\end{equation}
\end{definition}
Results in contraction theory can be applied to a quite general class of subsets $\mathcal{C}\subseteq\mathbb{R}^n$, known as $K$-reachable subsets \citep{russo2010global}.
\begin{definition}
Let $K>0$ be any positive real number. A subset $\mathcal{C}\subseteq\mathbb{R}^n$ is \emph{$K$-reachable} if, for any two points $x_0$ and $y_0$ in $\mathcal{C}$ there is some continuously differentiable curve $\gamma:[0,1]\rightarrow\mathcal{C}$ such that $\gamma(0)=x_0$, $\gamma(1)=y_0$ and $\lvert \gamma'(r)\rvert \leq K\lvert y_0-x_0\rvert,\; \forall\, r$.
\end{definition}
For convex sets $\mathcal{C}$, we may pick $\gamma(r)=x_0+r(y_0-x_0)$, so $\gamma'(r)=y_0-x_0$ and we can take $K=1$. Thus, convex sets are 1-reachable, and it is easy to show that the converse holds.

The main result of contraction theory for smooth systems is as follows \cite{lohmiller1998contraction,russo2010global}.
\begin{theorem}
\label{thm:contraction}
Let $\mathcal{C}\subseteq U$ be a forward-invariant $K$-reachable set. If there exists some norm in $\mathcal{C}$, with associated matrix measure $\mu$, such that, for some constant $c>0$ (the \emph{contraction rate})
\begin{equation}
\mu\left(\frac{\partial f}{\partial x}(t,x)\right)\leq -c \quad \quad \forall x\in\mathcal{C},\;\forall t\geq t_0,
\end{equation}
that is, the vector field \eqref{eq:dynamical_sys} is \emph{contracting} in $\mathcal{C}$, then system \eqref{eq:dynamical_sys} is incrementally exponentially stable in $\mathcal{C}$ with convergence rate $c$.
\end{theorem}
As a result, if a system is contracting in a (bounded) forward invariant subset then it converges towards an equilibrium point therein \citep{russo2010global,lohmiller1998contraction}.

In this paper we analyse contraction properties of dynamical systems based on norms and matrix measures. Other more general definitions exist in the literature, for example results based on Riemannian metrics \citep{lohmiller1998contraction} and Finsler-Lyapunov functions \citep{forni2014differential}. The relations between these three definitions and the definition of convergence \citep{pavlov2004convergent} have been investigated in \citep{forni2014differential}.
\subsection{Filippov systems}
Switched (or bimodal) Filippov systems are dynamical systems $\dot{x}=f(x)$ where $f(x)$ is a piecewise continuous vector field having a codimension one submanifold $\Sigma$ as its discontinuity set and defined as 
\begin{equation}
\label{eq:filippov_bimodal}
f(x)=
\begin{cases}
f^+(x) \quad \mbox{if } x\in\mathcal{S}^+ \\
f^-(x) \quad \mbox{if } x\in\mathcal{S}^-
\end{cases}
\end{equation}
where $f^+,f^-\in{C}^1(U,\mathbb{R}^n)$. The vector field $f(x)$ can be multivalued at the points of $\Sigma$.
The submanifold $\Sigma$ is defined as the zero set of a smooth function $H:\,U\rightarrow\mathbb{R}$, that is
\begin{equation}
\label{eq:switchingmanifold}
\Sigma:=\{x\in U : H(x)=0\} 
\end{equation}
where $0\in\mathbb{R}$ is a regular value of $H$, i.e. $\forall x\in\Sigma$
$$
\nabla H(x)=\left[ \frac{\partial H(x)}{\partial x_1}\;\; \dots \;\;\frac{\partial H(x)}{\partial x_n} \right] \neq 0.
$$
$\Sigma$ is called the {\it switching manifold}. It divides $U$ in two disjoint regions, $\mathcal{S}^+:=\{x\in U : H(x)>0\}$ and ${\mathcal{S}^-:=\{x\in U : H(x)<0\}}$.
We distinguish the following regions on $\Sigma$:
\begin{enumerate}
\item
The \emph{crossing region} is $\Sigma_c:=\{x\in\Sigma: \mathcal{L}_{f^+}H(x) \cdot \mathcal{L}_{f^-}H(x) >0 \}$;
\item
The \emph{sliding region} is $\Sigma_s:=\{x\in\Sigma: \mathcal{L}_{f^+}H(x)<0,\; \mathcal{L}_{f^-}H(x) >0 \}$;
\item
The \emph{escaping region} is $\Sigma_e:=\{x\in\Sigma: \mathcal{L}_{f^+}H(x)>0,\; \mathcal{L}_{f^-}H(x) <0 \}$;
\end{enumerate}
where $\mathcal{L}_{f\pm}H(x):=\nabla H(x)\, f^\pm(x)$ is the \emph{Lie derivative} of $H(x)$ with respect to the vector field $f^\pm(x)$, that is the component of $f^\pm(x)$ normal to the switching manifold at the point $x$. In the sliding region we adopt the widely used Filippov convention \citep{filippov1988differential}. We define a \emph{sliding vector field} $f^s$, which is the convex combination of $f^+$ and $f^-$ that is tangent to $\Sigma$, given for $x\in\Sigma_s$ by 
\begin{equation}
\label{eq:sliding}
f^s(x)=(1-\lambda)f^+(x)+\lambda f^-(x), \qquad \lambda\in[0,1]
\end{equation}
with $\lambda$ such that $\nabla H(x)\, f^s(x)=0$. 
\begin{remark}
In the following we assume that solutions of systems \eqref{eq:filippov_bimodal} and \eqref{eq:sliding} are defined in the sense of  Filippov and  that for \eqref{eq:filippov_bimodal} \emph{right uniqueness} \citep[pag. 106]{filippov1988differential} holds in $U$. Therefore, the escaping region is excluded from our analysis. 
\end{remark}
%
%
There are a few results on the incremental stability of piecewise smooth systems; notably for piecewise affine (PWA) systems and piecewise smooth continuous (PWSC) systems.
%
\begin{definition}[PWA systems]
A bimodal PWA system is a system of the form
\begin{equation}
\label{eq:pwa_sys}
\dot{x}=
\begin{cases}
A_1x+b_1+Bu \quad \mbox{if } h^Tx>0\\
A_2x+b_2+Bu \quad \mbox{if } h^Tx<0
\end{cases}
\end{equation}
where $x,\,h\in\mathbb{R}^n$, $u\in\mathbb{R}^m$, and $A_i\in\mathbb{R}^{n\times n}$, $B\in\mathbb{R}^{n\times m}$, $b_i\in\mathbb{R}^n$, $i=1,2$, are constant matrices and vectors, respectively.
\end{definition}
\begin{theorem}
\label{thm:pwa_pavlov}
\citep{pavlov2007convergence} System \eqref{eq:pwa_sys} is incrementally exponentially stable if there exist a positive definite matrix $P=P^T>0$, a number $\gamma\in\{0,1\}$ and a vector $g\in\mathbb{R}^n$ such that
\begin{enumerate}
\item
\label{eq:thm:pavlov:1}
$PA_i+A_i^TP<0, \quad i=1,2$,
\item
\label{eq:thm:pavlov:2}
$\Delta A=gh^T$,
\item
\label{eq:thm:pavlov:3}
$P\Delta b=-\gamma h$,
\end{enumerate}
where $\Delta A:=A_1-A_2$ and $\Delta b=b_1-b_2$.
\end{theorem}
\begin{remark}
The first condition requires the existence of a common Lyapunov function $V(x)=x^TPx$ for the two modes. The second condition assumes that the linear part of the two modes is continuous on the switching plane. There are two cases in the third condition  \citep[see][Remark 4]{pavlov2007convergence}. For $\gamma=0$, the PWA system \eqref{eq:pwa_sys} is continuous. For $\gamma=1$, the discontinuity is due only to the $b_i$ and, together with the first condition, implies that the two modes of the PWA system \eqref{eq:pwa_sys} are simultaneously strictly passive. 
\end{remark}
The original theorem in \citep[Theorem 2]{pavlov2007convergence}  is stated in terms of convergence instead of incremental stability. These two notions are proved to be equivalent on a compact set in \citep{ruffer2013convergent}.
%
%
%
\begin{definition}[PWSC systems]
\label{def:PWSC}
The piecewise smooth system \eqref{eq:pws} is said to be continuous (PWSC) if the following conditions hold:
\begin{enumerate}
\item
it is continuous for all $x\in\mathbb{R}^n$ and for all $t\geq t_0$
\item
the function $f_i(t,x)$ is continuously differentiable for all $x\in\mathcal{S}_i$, for all $t\geq t_0$ and for all $i$. Furthermore the Jacobian $\frac{\partial f_i}{\partial x}(t,x)$ can be continuously extended on the boundary $\partial S_i$.
\end{enumerate}
\end{definition}
\begin{theorem}
\label{thm:contracting_PWSC}
\citep{di2014contraction} Let $\mathcal{C}\subseteq U$ be a forward-invariant $K$-reachable set. Consider a PWSC system such that it fulfills conditions for the existence and uniqueness of a Carath\'eodory solution. If there exists a unique matrix measure such that for some positive constants $c_i$
\begin{equation*}
\mu\left( \frac{\partial f_i}{\partial x}(t,x) \right)\leq -c_i,
\end{equation*}
for all $x\in\bar{\mathcal{S}}_i$, for all $t\geq t_0$ and for all $i$, then the system is incrementally exponentially stable in $\mathcal{C}$ with convergence rate $c:=\min_i\, c_i$.
\end{theorem}
A similar result using Euclidean norms was previously presented in \citep[Theorem 2.33]{pavlov2006uniform} in terms of convergent systems. An extension of Theorem \ref{thm:contracting_PWSC} to the case where multiple norms are used was presented in \citep{lu2015switching,lu2015contraction}.
\subsection{Regularization}
\begin{figure}[!t]
\begin{center}
\includegraphics[width=0.9\linewidth]{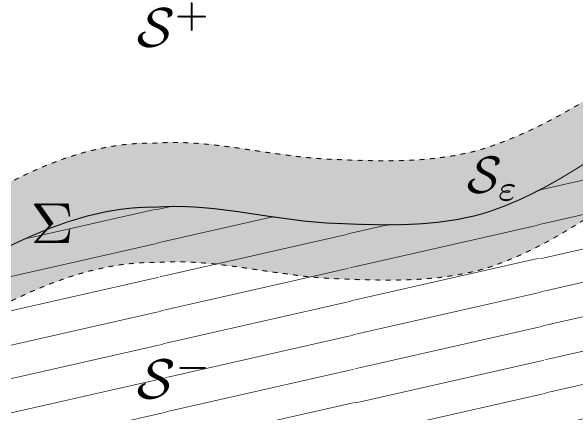}
\caption{Regions of state space: the switching manifold $\Sigma:=\{x\in U : H(x)=0\}$, 
$\mathcal{S}^+:=\{x\in U : H(x)>0\}$, $\mathcal{S}^-:=\{x\in U : H(x)<0\}$ (hatched zone) and $\mathcal{S}_{\varepsilon}:=\{x\in U : -\varepsilon <H(x)<\varepsilon \}$ (grey zone).}
\label{fig:regions}
\end{center}
\end{figure}

Our approach to contraction analysis of Filippov systems is via regularization. There are several ways to regularize system \eqref{eq:filippov_bimodal}. We shall adopt the method due to Sotomayor and Teixeira \citep{sotomayor1996regularization}, where a smooth approximation of the discontinuous vector field is obtained by means of a transition function.
\begin{definition}
A PWSC function $\varphi:\,\mathbb{R}\rightarrow\mathbb{R}$ is a \emph{transition function} if
\begin{equation}
\label{eq:transition_func}
 \varphi(s)=
 \begin{cases}
 1 & \text{if}\quad s\ge 1,\\
 \in (-1,1)& \text{if}\quad s\in (-1,1),\\
-1 & \text{if}\quad s\le -1,
 \end{cases}
 \end{equation}
and $\varphi'(s)>0$ within $s\in (-1,1)$.
\end{definition}
\begin{definition}
The $\varphi$\emph{-regularization} of a bimodal Filippov system \eqref{eq:filippov_bimodal} is the one-parameter family of PWSC functions  $f_\varepsilon: U \rightarrow \mathbb{R}^n$ given for $\varepsilon>0$ by
\begin{equation}
\label{eq:regularized_sys}
f_\varepsilon(x)\! = \! \frac{1}{2} \! \left[ 1 \! + \! \varphi\! \left(\! \tfrac{H(x)}{\varepsilon}\! \right)\right]\! f^+(x) \!+\!
\frac{1}{2} \! \left[ 1\! - \! \varphi\! \left(\! \tfrac{H(x)}{\varepsilon}\! \right)\right]\! f^-(x)
\end{equation}
\end{definition}
The \emph{region of regularization} where this process occurs is
\begin{equation*}
\label{eq:regularizationregion}
\mathcal{S}_{\varepsilon}:=\{x\in U : -\varepsilon <H(x)<\varepsilon \}.
\end{equation*}
Note that outside $\mathcal{S}_\varepsilon$ the regularized vector field $f_\varepsilon$ coincides with the PWS dynamics, i.e.
\begin{equation}
\label{eq:f_eps_outside}
f_\varepsilon(x)=
\begin{cases}
f^+(x) \quad \mbox{if } x\in\mathcal{S}^+\setminus \mathcal{S}_\varepsilon \\
f^-(x) \quad \mbox{if } x\in\mathcal{S}^- \setminus \mathcal{S}_\varepsilon 
\end{cases}
\end{equation}
A graphical representation of the different regions of the state space of the regularized vector field $f_\varepsilon$ is depicted in Figure~\ref{fig:regions}.

Sotomayor and Teixeira showed that the sliding vector field $f^s$ can be obtained as a limit of the regularized system in the plane. For $\mathbb{R}^n$, a similar result was given in \citep[Theorem 1.1]{llibre_sliding_2008}. Here we recover their results directly via the theory of slow-fast systems \citep{kuehn2015book} as follows.
\begin{lem}
\label{thm:regularization}
Consider $f$ in \eqref{eq:filippov_bimodal} with $0\in U$ and its regularization $f_\varepsilon$ in \eqref{eq:regularized_sys}. If for any $x\in\Sigma$ we have that $\mathcal{L}_{f^+}H(x)\neq 0$ or $\mathcal{L}_{f^-}H(x)\neq 0$ then there exists a singular perturbation problem such that fixed points of the boundary-layer model are critical manifolds, on which the motion of the slow variables is described by the reduced problem, which coincides with the sliding equations \eqref{eq:sliding}.\\
Furthermore, denoting by $x_\varepsilon(t)$ a solution of the regularized system and by $x(t)$ a solution of the discontinuous system with the same initial conditions $x_0$, then
\begin{equation*}
\lvert x_\varepsilon(t)-x(t)\rvert=O(\varepsilon)
\end{equation*}
uniformly for all $t\geq t_0$ and for all $x_0\in U$.
\end{lem}
\begin{pf*}{Proof.}
For the sake of clarity, we assume without loss of generality that $\Sigma$ can be represented, through a local change of coordinates around a point $x\in\Sigma$, by the function ${H(x)=x_1}$. We use the same notation for both coordinates. Hence our regularized system \eqref{eq:regularized_sys} becomes
\begin{equation}
\label{eq:regularized_sys_2}
\dot{x}=\dfrac{1}{2}\left[ 1+\varphi \left( \tfrac{x_1}{\varepsilon} \right)\right] f^+(x) +
\frac{1}{2}\left[ 1-\varphi \left( \tfrac{x_1}{\varepsilon} \right)\right] f^-(x)
\end{equation}
We now write \eqref{eq:regularized_sys_2} as a slow-fast system. Let $\hat x_1 = x_1/\varepsilon$, so that the region of regularization becomes $\hat x_1 \in (-1,1)$, and $\hat{x}_i=x_i$ for $i=2, \dots, n$. Then \eqref{eq:regularized_sys_2} can be written as
\begin{equation}
\label{eq:slow_system}
\begin{split}
\varepsilon\dot{\hat x}_1 & =  \frac{1}{2} \big[ 1+\varphi (\hat x_1) \big] f^+_1(\hat x) +
\frac{1}{2}\big[ 1-\varphi (\hat x_1)\big] f^-_1(\hat x), \\
\dot{\hat x}_i & =  \frac{1}{2} \big[ 1+\varphi (\hat x_1)\big] f^+_i(\hat x) +
\frac{1}{2}\big[ 1-\varphi (\hat x_1)\big] f^-_i(\hat x), 
\end{split}
\end{equation}
for $i=2, \dots, n$, where $\hat x = (\hat x_1, \hat x_2, \dots, \hat x_n)$. The variable $\hat x_1$ is the {\it fast variable} and the variables $\hat x_i$ for $i=2,\dots,n$ are the {\it slow variables}.
When $\varepsilon = 0$, we have
\begin{equation}
\label{eq:reduced_problem}
\begin{split}
0 & = \frac{1}{2} \big[ 1+\varphi (\hat x_1)\big]\, f^+_1(\hat x) +
\frac{1}{2}\big[ 1-\varphi (\hat x_1)\big]\, f^-_1(\hat x),\\
\dot{\hat x}_i & =  \frac{1}{2} \big[ 1+\varphi (\hat x_1)\big]\, f^+_i(\hat x) +
\frac{1}{2}\big[ 1-\varphi (\hat x_1)\big]\, f^-_i(\hat x),
\end{split}
\end{equation}
for $i=2, \dots, n$, obtaining the so-called \emph{reduced problem}. From the hypotheses we know that $f_1^+(\hat{x})\neq 0$ or $f_1^-(\hat{x})\neq 0$, hence we can solve for $\varphi$ from the first equation
\begin{equation}
\varphi(\hat{x}_1)=-\frac{f_1^+(\hat{x})+f_1^-(\hat{x})}{f_1^+(\hat{x})-f_1^-(\hat{x})},
\end{equation}
that substituted into the second equation in \eqref{eq:reduced_problem} gives
\begin{equation}
\label{eq:reduced_problem_2}
\dot{\hat{x}}_i=\frac{f_1^+(\hat{x})f_i^-(\hat{x})-f_i^+(\hat{x})f_1^-(\hat{x})}{f_1^+(\hat{x})-f_1^-(\hat{x})}, \quad i=2, \dots, n.
\end{equation}
If we now rescale time $\tau = t/\varepsilon$ and write $()^{'}=d/d\tau$, then \eqref{eq:slow_system} becomes
\begin{equation}
\label{eq:fast_system}
\begin{split}
{\hat x}_1^{'} & =  \frac{1}{2} \big[ 1+\varphi (\hat x_1)\big]\, f^+_1(\hat x) +
\frac{1}{2}\big[ 1-\varphi (\hat x_1)\big]\, f^-_1(\hat x),\\
{\hat x}_i^{'} & =  \frac{\varepsilon}{2} \big[ 1+\varphi (\hat x_1)\big]\, f^+_i(\hat x) +
\frac{\varepsilon}{2}\big[ 1-\varphi (\hat x_1)\big]\, f^-_i(\hat x), 
\end{split}
\end{equation}
for $i=2, \dots, n$. The limit $\varepsilon=0$ of \eqref{eq:fast_system}
\begin{equation*}
\label{eq:layer_problem}
\begin{split}
{\hat x}_1^{'} & =  \frac{1}{2} \big[ 1+\varphi (\hat x_1)\big]\, f^+_1(\hat x) +
\frac{1}{2}\big[ 1-\varphi (\hat x_1)\big]\, f^-_1(\hat x), \\
{\hat x}_i^{'} & = 0, \quad i=2, \dots,n, 
\end{split}
\end{equation*}
is called the {\it boundary-layer model}. Its fixed points can be obtained by applying the Implicit Function Theorem to $\hat{x}_1^{'}=0$, that gives $\hat{x}_1=h(x_2,\dots,x_n)$, since $\varphi'(\hat{x}_1)>0$ for $\hat{x}_1\in (-1,1)$ by definition. This in turn implies that $x_1=\varepsilon\, h(x_2,\dots,x_n)$.

It now follows directly that the flow of the reduced problem on critical manifolds of the boundary-layer problem coincides with that of the sliding vector field $f^s$ as in \eqref{eq:sliding} when the same change of coordinates as in the beginning is considered, i.e. such that $\nabla H=[1\;0\;\dots\;0]$. In fact, after some algebra we get
\begin{equation*}
f^s(x)=
\left[
0,\; \frac{f_1^+ f_2^- -f_2^+f_1^-}{f_1^+-f_1^-}, \; \dots \;,\; \frac{f_1^+ f_n^- -f_n^+f_1^-}{f_1^+-f_1^-}
\right]^T
\end{equation*}
that coincides with \eqref{eq:reduced_problem_2}.

Furthermore, it is a well known fact in singular perturbation problems \citep[Theorem 11.1]{khalil1996nonlinear} that, starting from the same initial conditions, the error between solutions  $\hat{x}(t)$ of the slow system \eqref{eq:slow_system} and solutions of its reduced problem (that, as said, coincide with solutions $x_s(t)$ of the sliding vector field) is $O(\varepsilon)$ after some $t_b>t_0$ when the fast variable $\hat{x}_1$ has reached a $O(\varepsilon)$ neighborhood of the slow manifold, i.e. $\lvert \hat{x}(t)-x_s(t) \rvert =O(\varepsilon), \;\, \forall t\geq t_b$. However, in our case the singular perturbation problem is defined only in $\mathcal{S}_\varepsilon$ where any point therein is distant from the slow manifold at most $2\varepsilon$, therefore the previous estimate is defined uniformly for all $t\geq t_0$ and in any norm due to their equivalence in finite dimensional spaces.
On the other hand, from \eqref{eq:f_eps_outside} outside $\mathcal{S}_\varepsilon$ the regularized vector field is equal to the discontinuous vector field and therefore the error between their solutions is uniformly 0.
\end{pf*}
\section{Contracting Filippov systems}
\label{sec:filippov}
In this section we present our two main results, Theorems \ref{thm:contraction_filippov} and \ref{thm:contracting_regularized}, for switched Filippov systems. Theorem \ref{thm:contraction_filippov}, using Lemma \ref{thm:regularization}, shows that if the regularized system $\dot{x}=f_\varepsilon(x)$ is incrementally exponentially stable so it is the Filippov system from which it is derived. 
Theorem \ref{thm:contracting_regularized} then gives sufficient conditions for the discontinuous vector field to be incrementally exponentially stable.
\begin{theorem}
\label{thm:contraction_filippov}
Let $\mathcal{C}\subseteq U$ be a forward-invariant $K$-reachable set. If there exists a positive constant $\bar{\varepsilon}<1$ such that for all $\varepsilon<\bar{\varepsilon}$ the regularized vector field $f_\varepsilon$ \eqref{eq:regularized_sys} is incrementally exponentially stable in $\mathcal{C}$ with convergence rate $c$, then in the limit for $\varepsilon\rightarrow 0^+$ any two solutions $x(t)=\psi(t,t_0,x_0)$ and $y(t)=\psi(t,t_0,y_0)$, with $x_0,y_0\in\mathcal{C}$, of the bimodal Filippov system \eqref{eq:filippov_bimodal} converge towards each other in $\mathcal{C}$, i.e.
\begin{equation}
\label{eq:thm:ies_contraction_filippov}
\lvert x(t)-y(t) \rvert \leq K\; e^{-c(t-t_0)}\lvert x_0-y_0\rvert, \quad \forall t\geq t_0.
\end{equation}
\end{theorem}
\begin{pf*}{Proof.}
From Lemma \ref{thm:regularization} we know that the error between any two solutions $x_\varepsilon(t)$ and $y_\varepsilon(t)$ of the regularized vector field $f_\varepsilon$ and their respective limit solutions $x(t)$ and $y(t)$ of the discontinuous system is $O(\varepsilon)$, i.e. $|x_\varepsilon(t)-x(t)|=O(\varepsilon)$ and $|y_\varepsilon(t)-y(t)|=O(\varepsilon)$, $\forall t\geq t_0$. Therefore, from the hypothesis of $f_\varepsilon$ being incrementally exponentially stable, \eqref{eq:ies} holds and applying the triangular inequality of norms we have
\begin{equation*}
\begin{split}
|x(t)-y(t)| \leq & \; |x(t)-x_\varepsilon(t)| + |x_\varepsilon(t)-y(t)|\\
\leq & \; |x(t)-x_\varepsilon(t)| + |x_\varepsilon(t)-y_\varepsilon(t)| \\
& + |y_\varepsilon(t)-y(t)|\\
\leq & \; K\;  e^{-c(t-t_0)} |x_\varepsilon(t_0)-y_\varepsilon(t_0)|\, + 2\, O(\varepsilon)
\end{split}
\end{equation*}
for $x_\varepsilon(t_0),y_\varepsilon(t_0)\in\mathcal{C}$ and for every $t\geq t_0$. 
The theorem is then proved by taking the limit for $\varepsilon\rightarrow 0^+$.
\end{pf*}
If the chosen transition function $\varphi$ is a $C^1(\mathbb{R})$ function, then the regularized vector field $f_\varepsilon$ is $C^1(U,\mathbb{R}^n)$ and Theorem \ref{thm:contraction} can be directly applied to study its incremental stability. On the other hand, if the transition function is not $C^1$ but it is at least a PWSC function as in Definition \ref{def:PWSC}, with $\mathcal{S}_1=(-\infty,-1)$, $\mathcal{S}_2=(-1,1)$ and $\mathcal{S}_3=(1,+\infty)$, then the regularized vector field $f_\varepsilon$ is itself a PWSC vector field and Theorem \ref{thm:contracting_PWSC} applies. This is the case for $\varphi(s)=\mathrm{sat}(s)$. This function is $C^0(\mathbb{R})$ but its restrictions to each subsets $\mathcal{S}_1$, $\mathcal{S}_2$ and $\mathcal{S}_3$ are smooth functions. We will use it as an example in the sequel.

Before presenting our next theorem, we first introduce the following lemma.
\begin{lem}
\label{lemma:jacobian}
The Jacobian matrix of the regularized vector field \eqref{eq:regularized_sys} is
\begin{equation}
\label{eq:jacobian_eps}
\begin{split}
\frac{\partial f_\varepsilon}{\partial x}(x)=&
\;\alpha(x)\, \frac{\partial f^+}{\partial x} (x) +
\beta(x)\, \frac{\partial f^-}{\partial x} (x) \\
&+\gamma(x)\,\Big[f^+(x)-f^-(x) \Big]\, \nabla H(x)
\end{split}
\end{equation}
where
\begin{eqnarray*}
\alpha(x):= \frac{1}{2}\left[ 1+ \varphi\left( \frac{H(x)}{\varepsilon} \right)\right]\\
\beta(x):= \frac{1}{2}\left[ 1- \varphi\left( \frac{H(x)}{\varepsilon} \right)\right]\\
\gamma(x):= \frac{1}{2\varepsilon}\,\varphi'\left( \frac{H(x)}{\varepsilon} \right)
\end{eqnarray*}
and $\alpha(x)\in [0,1]$, $\beta(x)\in[0,1]$ and $\gamma(x)\geq 0$, $\forall x\in U,\; \forall\varepsilon>0$. Note that for any transition functions $\alpha(x)+\beta(x)=1$, for all $x$.
\end{lem}
\begin{pf*}{Proof.}
The regularized vector field $f_\varepsilon$ can be rewritten as
$$
f_\varepsilon(x)=\alpha(x)f^+(x)+\beta(x) f^-(x)
$$
therefore, taking the derivative with respect to $x$, we obtain
\begin{equation}
\label{eq:lemma:jacobian}
\begin{split}
\frac{\partial f_\varepsilon}{\partial x}(x)= &\; \alpha(x)\frac{\partial f^+}{\partial x}(x)+\beta(x)\frac{\partial f^-}{\partial x}(x)\\
& +f^+(x)\frac{\partial\alpha}{\partial x}(x)+f^-(x)\frac{\partial\beta}{\partial x}(x).
\end{split}
\end{equation}
Observing that
\begin{equation*}
\begin{split}
\frac{\partial \alpha}{\partial x}(x)= & \frac{1}{2}\;\frac{\partial \varphi}{\partial s}\left( \frac{H(x)}{\varepsilon}\right) \, \frac{\partial}{\partial x}\left[ \frac{H(x)}{\varepsilon} \right]\\
=& \frac{1}{2\varepsilon}\; \varphi'\left(\frac{H(x)}{\varepsilon}\right) \nabla H(x)=\gamma(x)\,\nabla H(x)
\end{split}
\end{equation*}
and
$$
\frac{\partial \beta}{\partial x}(x)=-\frac{\partial \alpha}{\partial x}(x),
$$
replacing them into \eqref{eq:lemma:jacobian}, we finally obtain \eqref{eq:jacobian_eps}.
\end{pf*}
Note that if $\varphi$ is PWSC then the Jacobian matrix \eqref{eq:jacobian_eps} is a discontinuous function but its restriction to $\mathcal{S}_\varepsilon$ is continuous.
\begin{theorem}
\label{thm:contracting_regularized}
Let $\mathcal{C}\subseteq U$ be a forward-invariant $K$-reachable set. A bimodal Filippov system \eqref{eq:filippov_bimodal} is incrementally exponentially stable in $\mathcal{C}$ with convergence rate $c:=\min\,\{c_1,c_2\}$ if there exists some norm in $\mathcal{C}$, with associated matrix measure $\mu$, such that for some positive constants $c_1,c_2$
\begin{eqnarray}
\label{eq:thm:condition1}
\mu\left( \frac{\partial f^+}{\partial x}(x)\right) \leq -c_1,\quad \forall x \in \bar{\mathcal{S}}^+\\
\label{eq:thm:condition2}
\mu\left( \frac{\partial f^-}{\partial x}(x)\right) \leq -c_2, \quad \forall x \in \bar{\mathcal{S}}^-\\ 
\label{eq:thm:condition3}
\mu\left( \Big[ f^+(x)-f^-(x)\Big] \, \nabla H(x) \right) = 0, \quad \forall x \in \Sigma.
\end{eqnarray}
\end{theorem}
\begin{pf*}{Proof.}
The transition function $\varphi$ is a PWSC function hence the resulting regularized vector field $f_\varepsilon$ is also PWSC, i.e. it is continuous in all $U$ and such that its restrictions to the subsets $\bar{\mathcal{S}}^+\setminus\mathcal{S}_\varepsilon$, $\bar{\mathcal{S}}^-\setminus\mathcal{S}_\varepsilon$ and $\bar{\mathcal{S}}_\varepsilon$ are continuously differentiable. Therefore Theorem \ref{thm:contracting_PWSC} can be directly applied and we have that $f_\varepsilon$ is contracting in $\mathcal{C}$ if there exist positive constants $c_1,c_2,c_3$ such that
\begin{eqnarray}
\label{eq:contracting_mode1}
\mu\left( \frac{\partial f^+}{\partial x}(x)\right) &\leq -c_1,\quad &\forall x \in \bar{\mathcal{S}}^+\setminus\mathcal{S}_\varepsilon\\
\label{eq:contracting_mode2}
\mu\left( \frac{\partial f^-}{\partial x}(x)\right) &\leq -c_2, \quad &\forall x \in \bar{\mathcal{S}}^- \setminus \mathcal{S}_\varepsilon\\
\label{eq:contracting_regularized} 
\mu \left(\frac{\partial f_\varepsilon}{\partial x}(x)\right) &\leq -c_3, \quad &\forall x \in \bar{\mathcal{S}}_\varepsilon.
\end{eqnarray}
Thus, by Lemma \ref{lemma:jacobian}, substituting \eqref{eq:jacobian_eps} into \eqref{eq:contracting_regularized} and using the subadditivity and positive homogeneity properties of the matrix measures, we obtain
\begin{equation}
\label{eq:thm:subadditivity}
\begin{split}
\mu  \left( \frac{\partial f_\varepsilon}{\partial x} (x)\right)\leq & \;\alpha(x) \,\mu\left( \frac{\partial f^+}{\partial x} (x)\right)+ \beta(x) \,\mu\left( \frac{\partial f^-}{\partial x} (x)\right)\\
& +  \gamma(x) \,\mu\left( \Big[f^+(x)-f^-(x) \Big]\, \nabla H(x) \right)
\end{split}
\end{equation}
Therefore, conditions \eqref{eq:contracting_mode1}-\eqref{eq:contracting_regularized} are satisfied if
\begin{eqnarray}
\label{eq:thm:condition_reg_1}
\mu\left( \frac{\partial f^+}{\partial x}(x)\right) \leq -c_1,\quad \forall x \in \bar{\mathcal{S}}^+\cup \bar{\mathcal{S}}_\varepsilon\\
\label{eq:thm:condition_reg_2}
\mu\left( \frac{\partial f^-}{\partial x}(x)\right) \leq -c_2, \quad \forall x \in \bar{\mathcal{S}}^-\cup \bar{\mathcal{S}}_\varepsilon\\ 
\label{eq:thm:condition_reg_3}
\mu\left( \Big[ f^+(x)-f^-(x)\Big] \, \nabla H(x) \right) = 0, \qquad \forall x \in \bar{\mathcal{S}}_\varepsilon
\end{eqnarray}
and $c_3\geq \min\,\{c_1,c_2\}$. Finally, considering that $\bar{\mathcal{S}}_\varepsilon\rightarrow \Sigma$ in the limit for $\varepsilon\rightarrow 0^+$, we obtain conditions \eqref{eq:thm:condition1}-\eqref{eq:thm:condition3}. Therefore, by virtue of Theorem \ref{thm:contraction_filippov}, these conditions are sufficient for the bimodal Filippov system \eqref{eq:filippov_bimodal} to be incrementally exponentially stable.
\end{pf*}
\begin{remark}
If $\varphi$ is $C^1(\mathbb{R})$ it can be easily proved (by using Lemma \ref{lemma:jacobian} and the subadditivity property of matrix measures) that conditions \eqref{eq:thm:condition_reg_1}-\eqref{eq:thm:condition_reg_3} are sufficient for the measure of the Jacobian of $f_\varepsilon(x)$ to be negative definite over the entire region of interest.
\end{remark}
The first two conditions \eqref{eq:thm:condition1} and \eqref{eq:thm:condition2} in Theorem \ref{thm:contracting_regularized} guarantee that the regularized vector field $f_\varepsilon$ is contracting outside the region $\mathcal{S}_\varepsilon$, and therefore imply that any two trajectories in $\mathcal{C}\setminus\mathcal{S}_\varepsilon$ converge towards each other exponentially. Condition \eqref{eq:thm:condition3} assures that the third term in \eqref{eq:thm:subadditivity} does not diverge as $\varepsilon\to 0^+$ and therefore that negative definiteness of the measures of the Jacobian matrices of two modes, $f^+$ and $f^-$, is enough to guarantee incremental exponential stability of $f_\varepsilon$ inside $\mathcal{S}_\varepsilon$. 

Theorem \ref{thm:contracting_regularized} gives conditions in terms of a generic norm. When a specific norm is chosen, it is possible to further specify the conditions of Theorem \ref{thm:contracting_regularized}, as we now show.
\begin{proposition}
\label{thm:condition3}
Assume that through a local change of coordinates around a point $x\in\Sigma$ the switching manifold $\Sigma$ is represented by the function $H(x)=x_1$ and let $\Delta f(x)=f^+(x)-f^-(x)=[\Delta f_1(x)\; \dots \; \Delta f_n(x)]^T$. Let $D=\mathrm{diag}\{d_1,\,\dots,\, d_n\}$, with $d_i>0\; \forall i$, be a diagonal matrix and $P=Q^2$ be a positive definite matrix. Assuming that $\Delta f(x)\neq 0\;\;\forall x\in\Sigma$, then
\begin{enumerate}
\item
$\mu_{D,1}(\Delta f(x)\,\nabla H)= 0$ if and only if
\begin{equation*}\!\!\!\!
\begin{cases}
\Delta f_1(x)<0\\
\lvert \Delta f_1(x) \rvert \! \geq \! \lvert d_2 \Delta f_2(x) d_1^{-1} \rvert\! + \dots + \!\lvert d_n \Delta f_n(x) d_1^{-1}\rvert
\end{cases}
\end{equation*}
\item
$\mu_{Q,2}(\Delta f(x)\,\nabla H)= 0$ if and only if $P \Delta f(x)=-a\,\nabla H^T$, $a>0$.
\item
$\mu_{D,\infty}(\Delta f(x)\,\nabla H)= 0$ if and only if $\Delta f(x)$ and $\nabla H^T$ are antiparallel.
\end{enumerate}
\end{proposition}
\begin{pf*}{Proof.}
The matrix $(\Delta f(x)\,\nabla H)$ has rank equal to 1 and, since $\nabla H=[1\;\; 0\;\; \dots \;\; 0]$, it can be written as
\begin{equation*}
\label{eq:rank1_y}
\Delta f(x)\,\nabla H =
\begin{bmatrix}
\Delta f_1(x) & 0 & \dots & 0\\
\Delta f_2(x) & 0 & \dots & 0\\
\vdots & \vdots & \ddots & \vdots\\
\Delta f_n(x) & 0 & \dots & 0
\end{bmatrix}
\end{equation*}
\begin{enumerate}
\item
From \citep[Lemma 4]{vidyasagar1978matrix} we have
\begin{equation*}
\begin{split}
\mu_{D,1} & (\Delta f(x)\,\nabla H)=\\
& = \max\{ \Delta f_1(x)+|d_2 \Delta f_2(x) d_1^{-1}|+\dots+\\ 
& \qquad +|d_n \Delta f_n(x) d_1^{-1}|;0;\;\dots; \; 0 \}.
\end{split}
\end{equation*}
This measure is equal to zero if and only if
\begin{equation*}
\Delta f_1(x)+|d_2 \Delta f_2(x) d_1^{-1}|+\dots+|d_n \Delta f_n(x) d_1^{-1}|\leq 0.
\end{equation*}
\item
The proof for $\mu_{Q,2}$ comes from Lemma \ref{lemma:measure_rank1}.
\item
Again, from \citep[Lemma 4]{vidyasagar1978matrix} we have
\begin{equation*}
\begin{split}
\mu_{D,\infty} & (\Delta f(x)\,\nabla H)=\\
& = \max\{ \Delta f_1(x);\; |d_2\Delta f_2(x)d_1^{-1}|;\;\dots\;; \\
& \qquad |d_n\Delta f_n(x)d_1^{-1}| \}.
\end{split}
\end{equation*}
The above measure is equal to zero if and only if ${\Delta f_1(x)< 0}$ and $\Delta f_2(x)=\dots=\Delta f_n(x)=0$, that is if $\Delta f(x)$ is antiparallel to $\nabla H^T$.
\end{enumerate}
\end{pf*}
Hence, using the $\ell_1$-norm there always exist a matrix $D$ and a change of coordinates such that the condition holds assuming that the scalar product between $\nabla H$ and $\Delta f$ is negative, that is $\nabla H(x)\,\Delta f(x) <0,\;\forall x\in\Sigma$. Moreover, using the Euclidean norm a matrix $P$ such that the condition holds exists only if  $\nabla H(x)\,\Delta f(x) <0$, $\forall x\in\Sigma$, as proved next.
%
\begin{figure}[!t]
\begin{center}
\includegraphics[width=0.9\linewidth]{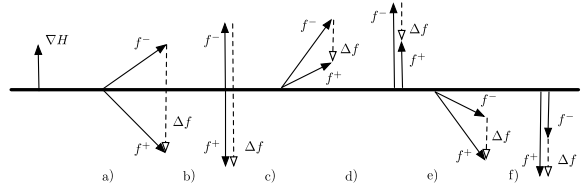}
\caption{Geometrical interpretation of condition \eqref{eq:thm:condition3} using Euclidean norm (with $Q=I$) and $\infty$-norm in $\mathbb{R}^2$. The horizontal line is $\Sigma$. Sliding is represented in a) and b), while crossing occurs in c), d), e), f). In all cases the difference vector field $\Delta f$ is antiparallel to $\nabla H$.}
\label{fig:geo_interp}
\end{center}
\end{figure}
\begin{proposition}
Assume that $\Delta f(\bar{x})\neq 0$ with $\bar{x}\in\Sigma$, then a Euclidean norm $|\cdot|_{Q,2}$, with $Q>0$, such that $\mu_{Q,2}(\Delta f(\bar{x})\,\nabla H(\bar{x}))= 0$ exists if and only if ${\nabla H(\bar{x})\, \Delta f(\bar{x})<0}$.
\end{proposition}
\begin{pf*}{Proof.}
Firstly, note that from Proposition \ref{thm:condition3} and from Lemma \ref{lemma:measure_rank1} we know that $\mu_{Q,2}( \Delta f(\bar{x})\,\nabla H(\bar{x})) = 0$ if and only if a matrix $P=Q^2$ exists such that $P\Delta f(\bar{x})=-a\,\nabla H(\bar{x})$, $a>0$. Now, from the definition of positive definite matrices, it follows that given the two nonzero vectors $\Delta f(\bar{x})$ and $\nabla H(\bar{x})$ such a positive definite matrix $P$ exists if and only if $-\nabla H(\bar{x})\,\Delta f(\bar{x})>0$, that is $\nabla H(\bar{x})\, \Delta f(\bar{x})<0$\footnote{Sufficiency follows directly from the definition of positive definiteness of the matrix $P$.}.
\end{pf*}
Furthermore, note that when $\Delta f(x)=0,\;\;\forall x\in\Sigma$, that is when the system is continuous on $\Sigma$ as in the case of PWSC systems,  we have that $\mu(\Delta f(x)\,\nabla H(x))=\mu(\varnothing)=0$. Therefore condition \eqref{eq:thm:condition3} is always satisfied and Theorem \ref{thm:contracting_regularized} coincides with Theorem \ref{thm:contracting_PWSC}.

In Figure \ref{fig:geo_interp} the geometrical interpretation of condition \eqref{eq:thm:condition3} in $\mathbb{R}^2$ is shown schematically when either the Euclidean norm (with $Q=I$) or the $\infty$-norm are used.
One significant advantage of our method is that it can deal with nonlinear PWS systems, as we shall now demonstrate. 
%
All simulations presented here were computed using the numerical solver in \citep{piiroinen2008event}.

\begin{figure}[!t]
\begin{center}
\includegraphics[width=0.9\linewidth]{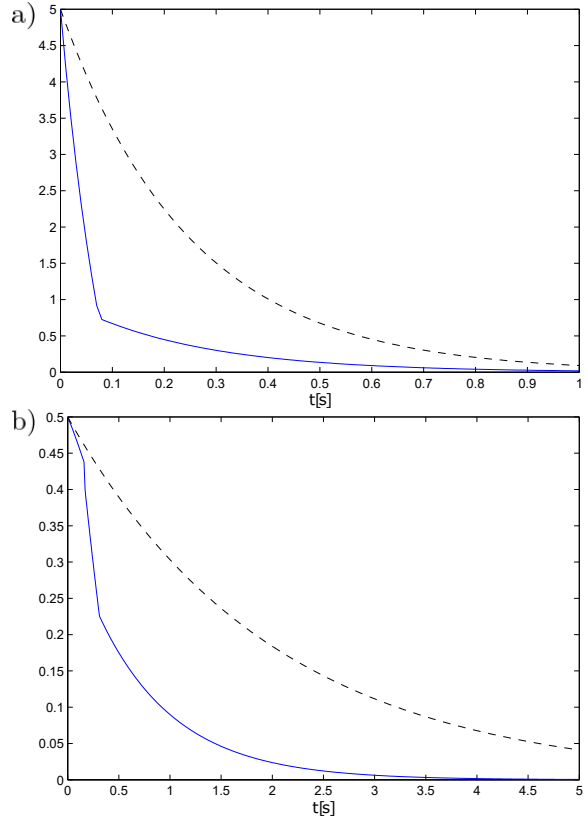}
\caption{ {Norm of the difference between two trajectories for (a) Example 1 and (b) Example 2. Initial conditions are respectively $x_0=[2\;\; 2]^T\in\mathcal{S}^+$, $y_0=[3\;\; -2]^T\in\mathcal{S}^-$ and $x_0=[0\;\; -1]^T\in\mathcal{S}^-$, $y_0=[0\;\; -0.5]^T\in\mathcal{S}^-$. The dashed lines represent the analytical estimates \eqref{eq:thm:ies_contraction_filippov} with (a) $c=4$ and (b) $c=1/2$, and $K=1$.}}
\label{fig:ex_pws}
\end{center}
\end{figure}

\paragraph*{Example 1}
Consider the PWS system \eqref{eq:filippov_bimodal} with
\begin{equation*}
\label{eq:pwsexample1}
f^+(x)\! =\!\!
\begin{bmatrix}
-4x_1\\
-9x_2-x_2^2-18
\end{bmatrix}\!,
\,
f^-(x)\! =\!\!
\begin{bmatrix}
-4x_1\\
-9x_2+x_2^2+18
\end{bmatrix}
\end{equation*}
and $H(x)=x_2$. We can easily check that all three conditions of Theorem \ref{thm:contracting_regularized} are satisfied in the $\ell_1$-norm. Indeed, for the first condition we have
\begin{equation*}
\mu_1\left(\frac{\partial f^+}{\partial x}(x)\right) = 
\max\{-4;\; -2x_2-9 \}=-4
\end{equation*}
because $-2x_2-9<-9,\;\; \forall x\in\mathcal{S}^+$. Similarly for the second condition we have
\begin{equation*}
\mu_1\left(\frac{\partial f^-}{\partial x}(x)\right) =\max\{-4;\; 2x_2-9 \}=-4
\end{equation*}
because $2x_2-9<-9,\;\; \forall x\in\mathcal{S}^-$. Finally, for the third condition we have
\begin{equation*}
\begin{split}
\mu_1 & \left(\Big[f^+(x)-f^-(x)\Big]\, \nabla H(x)\right)= \\
= & \mu_1\left(
\begin{bmatrix}
0 & 0\\
0 & -2x_2^2-36
\end{bmatrix}
\right)=\\
= & \max\{0;\; -2x_2^2-36\}=0, \qquad \forall x\in\Sigma.
\end{split}
\end{equation*}
Therefore the PWS system considered here is incrementally exponentially stable in all $\mathbb{R}^2$ with convergence rate $c=4$. In Figure \ref{fig:ex_pws}a we show numerical simulations which confirm the analytical estimation \eqref{eq:thm:ies_contraction_filippov}.
%
%
%
\paragraph*{Example 2}
Consider the PWS system \eqref{eq:filippov_bimodal} with
\begin{equation*}
\label{eq:pwsexample2}
f^+(x)\! =\!\!
\begin{bmatrix}
-2x_1-\dfrac{2}{9}x_2^2+\! 2\\[2ex]
x_1-x_2-3
\end{bmatrix}\! ,
f^-(x)\! =\!\!
\begin{bmatrix}
-2x_1+\dfrac{2}{9}x_2^2-\! 2\\[2ex]
x_1-x_2+3
\end{bmatrix}
\end{equation*}
and $H(x)=x_2$. For the first condition of Theorem \ref{thm:contracting_regularized} we have
\begin{equation*}
\begin{split}
\mu_1\left(\frac{\partial f^+}{\partial x}(x)\right) 
= & \max\left\{-1;\; -1+\frac{4}{9}\;|x_2| \right\}=\\
= & -1+\frac{4}{9}\;|x_2|
\end{split}
\end{equation*}
Therefore $f^+$ is contracting in the $\ell_1$-norm for $|x_2|<9/4$. If we want to guarantee a certain contraction rate  $c$ we need to consider the subset $|x_2|<9/4(1-c)$ instead. An identical result holds for $f^-$.
Finally, for the third condition of Theorem \ref{thm:contracting_regularized} we have
\begin{equation*}
\begin{split}
\mu_1 & \left(\Big[f^+(x)-f^-(x)\Big]\, \nabla H(x)\right)= \\
= & \mu_1\left(
\begin{bmatrix}
0 & -\dfrac{4}{9}x_2^2+4\\[2ex]
0 & -6
\end{bmatrix}
\right)=\\
=& \max\left\{0;\; -2+\frac{4}{9}x_2^2 \right\}=0
\end{split}
\end{equation*}
for all $x\in\Sigma$, that is $x_2=0$. 
We can conclude that the PWS system taken into example satisfies Theorem \ref{thm:contracting_regularized} in the subset $\mathcal{C}=\{x\in\mathbb{R}^2: \; |x_2|<9/8 \}$ and therefore it is incrementally exponentially stable  with convergence rate $c=1/2$ therein. This is confirmed by numerical simulations shown in Figure \ref{fig:ex_pws}b.
\section{Application to PWA systems}
\label{sec:examples}
%
%
\begin{proposition}
\label{thm:contracting_pwa}
The PWA system \eqref{eq:pwa_sys} is incrementally exponentially stable in a forward-invariant $K$-reachable set $\mathcal{C}\subseteq U$  with convergence rate $c:=\min\,\{c_1,c_2\}$ if there exists some norm in $\mathcal{C}$, with associated matrix measure $\mu$, such that for some positive constants $c_1,c_2$ and for all $x\in\Sigma$
\begin{eqnarray}
\label{eq:thm:pwa_cond1}
&\mu\left( A_1\right)& \leq -c_1\\
&\mu\left( A_2\right)& \leq -c_2\\
\label{eq:thm:pwa_cond3}
&\mu\left( \Delta Axh^T \right)& = 0\\
\label{eq:thm:pwa_cond4}
&\mu\left( \Delta bh^T \right)& = 0 
\end{eqnarray}
\end{proposition}
\begin{pf*}{Proof.}
The proof follows directly from Theorem \ref{thm:contracting_regularized} noting that $\frac{\partial f^+}{\partial x}=A_1$, $\frac{\partial f^-}{\partial x}=A_2$, $f^+(x)-f^-(x)=\Delta Ax+\Delta b$, and $\nabla H(x)=h^T$. Indeed 
\begin{equation*}
\begin{split}
\mu & \left( \Big[ f^+(x)-f^-(x)\Big] \, \nabla H(x) \right)=\\
= & \mu\left( [\Delta Ax+\Delta b] \, h^T \right) 
\leq \mu\left( \Delta Axh^T \right)+\mu\left( \Delta bh^T \right).
\end{split}
\end{equation*}
\end{pf*}

\begin{figure}[!t]
\begin{center}
\includegraphics[width=0.9\linewidth]{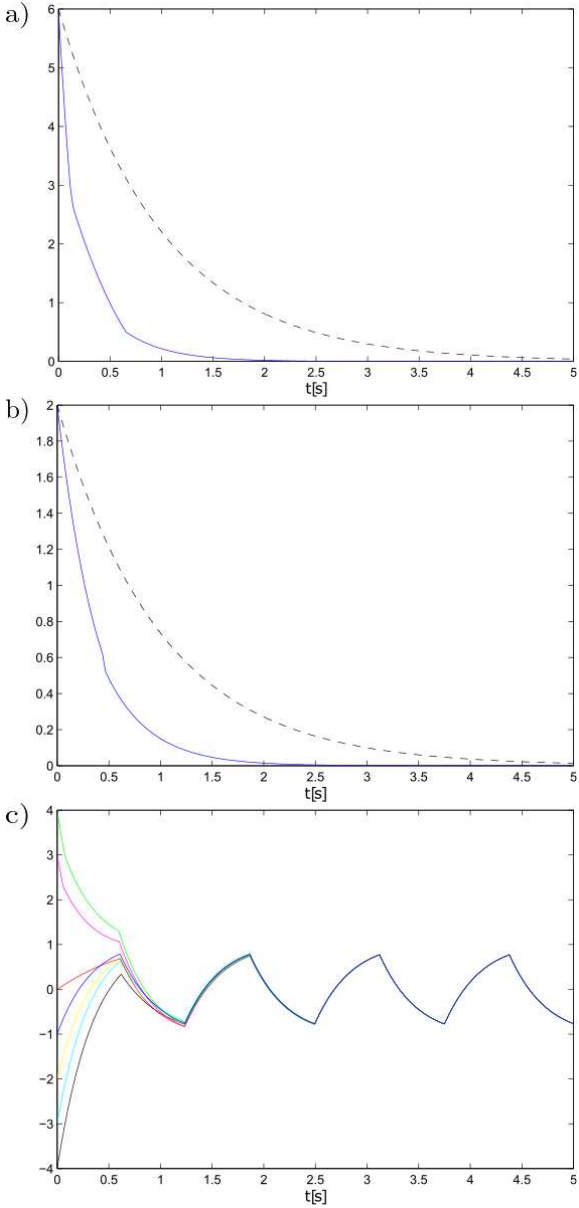}
\caption{Norm of the difference between two trajectories for (a) Example 3 and (c) Example 4. Initial conditions are respectively $x_0=[4\;\; 4]^T\in\mathcal{S}^+$, $y_0=[3\;\; -1]^T\in\mathcal{S}^-$ and $x_0=[2\;\; 2]^T\in\mathcal{S}^+$, $y_0=[2\;\; -2]^T\in\mathcal{S}^-$.  The dashed lines represent the analytical estimates \eqref{eq:thm:ies_contraction_filippov} with $K=1$ and $c=1$. Panel (b) depicts the time evolution of the state $x_2(t)$ of Example 3 from different initial conditions and with $u(t)=6\, \sin(2\pi\,t)$ set as a periodic input signal.}
\label{fig:ex_pwa}
\end{center}
\end{figure}

\begin{remark}
When Euclidean norms $|\cdot |_{Q,2}$ are used, with $Q=P^{1/2}$, the conditions of Proposition \ref{thm:contracting_pwa} become the same as those in Theorem \ref{thm:pwa_pavlov}. It is easy to show that the conditions of Theorem \ref{thm:pwa_pavlov} are sufficient for those of our Proposition to hold. In fact, from Theorem \ref{app:thm:weighted_mu}, condition \ref{eq:thm:pavlov:1} of Theorem \ref{thm:pwa_pavlov} on the matrices $A_1$ and $A_2$  implies that their measures $\mu_{Q,2}(A_1)$ and  $\mu_{Q,2}(A_2)$ are negative definite. Condition \ref{eq:thm:pavlov:2} of Theorem \ref{thm:pwa_pavlov} implies that in any norm
$
\mu\left( \Delta Axh^T\right)=\mu\left( g\,(h^Tx)\,h^T\right) =0
$
since $h^Tx=0$, ${\forall x\in\Sigma}$. 
Condition \ref{eq:thm:pavlov:3} of Theorem \ref{thm:pwa_pavlov} can be rewritten as $Q\Delta b=-Q^{-1}h$, therefore
$
\mu_{Q,2}\left( \Delta bh^T \right)=\mu_2\left(Q \Delta bh^T Q^{-1}\right) = \mu_2\left(-Q^{-1}h\; (Q^{-1}h)^T  \right)=0
$
for Lemma \ref{lemma:measure_rank1}, since vectors $Q^{-1}h$ and $-Q^{-1}h$ are antiparallel.
\end{remark}
\paragraph*{Example 3}
Consider a PWA system of the form \eqref{eq:pwa_sys} with
\begin{equation*}
\label{eq:pwaexample}
\begin{split}
&A_1=
\begin{bmatrix}
-2 & -1\\
1 & -3
\end{bmatrix},
\quad
b_1=
\begin{bmatrix}
-1\\
-3
\end{bmatrix},\\
&A_2=
\begin{bmatrix}
-2 & -1\\
1 & -4
\end{bmatrix},
\quad
b_2=
\begin{bmatrix}
2\\
4
\end{bmatrix},
\end{split}
\end{equation*}
and $B=[0\;\;1]^T$, $h=[0\;\;1]^T$. Using the $\ell_1$-norm the first two conditions of Proposition \ref{thm:contracting_pwa} are satisfied, in fact $\mu_1(A_1)=-1$ and $\mu_1(A_2)=-1$. The third condition is also satisfied since we have that
\begin{equation*}
\mu_1(\Delta Axh^T)
= \mu_1\left(
\begin{bmatrix}
0 & 0\\
0 & x_2
\end{bmatrix} 
\right)= x_2 =0, \qquad \forall x\in\Sigma.
\end{equation*}
Finally, the fourth condition is satisfied as it can be easily proved that $\mu_1(\Delta b h^T)=0$. Therefore, from Proposition \ref{thm:contracting_pwa}, the PWA system considered here is incrementally exponentially stable.
In Figure \ref{fig:ex_pwa}a we show numerical simulations of the norm of the difference between two trajectories for this PWA system. Similar qualitative behavior was observed for different choices of the initial conditions. The dashed line is the estimated exponential decay from \eqref{eq:thm:ies_contraction_filippov} with $c=1$ and $K=1$. It can be seen that as expected from the theoretical analysis
$
\lvert x(t)-y(t)\rvert_1 \leq e^{-t} \lvert x_0-y_0 \rvert_1, \; \forall t\geq 0.
$

The evolution of the system state $x_2(t)$ is reported in Figure \ref{fig:ex_pwa}b when the periodic signal $u(t)=6\,\sin(2\pi\,t)$ is chosen as a forcing input. As expected for contracting systems, all trajectories converge towards a unique periodic (non-smooth) solution with the same period of the excitation $u(t)$ (confirming the entrainment property of contracting systems reported e.g. in \citet{russo2010global}). 
\subsection{Relay feedback systems}
We present here a similar result for relay feedback systems.
\begin{proposition}
\label{thm:contracting_relay}
A relay feedback system of the form
\begin{equation}
\begin{split}
\label{eq:relay_sys}
\dot{x}=& \,Ax-b\;\mathrm{sgn}(y)\\
y=& \, c^Tx
\end{split}
\end{equation}
where $A\in\mathbb{R}^{n\times n}$, $b,\,c\in\mathbb{R}^n$, is incrementally exponentially stable in a forward-invariant $K$-reachable set $\mathcal{C}\subseteq U$ with convergence rate $\bar{c}$ if there exists some norm in $\mathcal{C}$, with associated matrix measure $\mu$, such that for some positive constant $\bar{c}$
\begin{eqnarray}
\label{eq:thm:relay_cond1}
\mu\left( A\right) \leq -\bar{c}\\
\label{eq:thm:relay_cond2}
\mu\left( -bc^T \right) =0.
\end{eqnarray}
\end{proposition}
\begin{pf*}{Proof.}
The proof follows observing that the relay feedback system is a PWA system of the form
\begin{equation*}
\dot{x}=
\begin{cases}
Ax-b \quad & \mathrm{if} \; c^Tx>0\\
Ax+b \quad & \mathrm{if} \; c^Tx<0
\end{cases}
\end{equation*}
with $A_1=A_2=A$, $\Delta A=\varnothing$, $\Delta b=-2b$ and $h=c$. Therefore applying Proposition \ref{thm:contracting_pwa} to it we get
$
\mu\left(\Delta Axh^T \right) + \mu\left( \Delta bh^T \right) =0 + \mu\left( -2bc^T \right) = 2\,\mu\left( -bc^T \right), 
$
and the assertion is proved.
\end{pf*}
\begin{remark}
It is known that if a smooth system is contracting in a forward invariant set then it must converge towards an equilibrium point, hence it cannot converge to a limit cycle. We show here that if conditions \eqref{eq:thm:relay_cond1} and \eqref{eq:thm:relay_cond2} hold then a planar relay feedback system \eqref{eq:relay_sys} cannot converge to a limit cycle either. In Euclidean norms condition \eqref{eq:thm:relay_cond1}  implies from Theorem \ref{app:thm:weighted_mu} that $A$ is Hurwitz, this in turn implies that its trace is negative, i.e. $\mathrm{tr}(A)<0$. Condition \eqref{eq:thm:relay_cond2} implies from Lemma \ref{lemma:measure_rank1} that $Pb=c$ where $P$ is a positive definite matrix, this means that $c^Tb=(Pb)^Tb=b^T Pb>0$ for any $b\neq 0$. The regularized vector field of \eqref{eq:relay_sys} is 
\begin{equation*}
f_\varepsilon(x)=Ax-b\,\varphi\left( \frac{c^Tx}{\varepsilon} \right)
\end{equation*}
If $\varphi\in C^1$ so it is also $f_\varepsilon$ and its divergence is
\begin{equation*}
\mathrm{div} (f_\varepsilon(x))=
\begin{cases}
\mathrm{tr}(A)-\frac{1}{\varepsilon}\varphi'\left( \frac{c^Tx}{\varepsilon}\right)c^T b, & \quad \mbox{if } x\in \mathcal{S}_\varepsilon\\
\mathrm{tr}(A), & \quad \mbox{if } x\notin \mathcal{S}_\varepsilon
\end{cases}
\end{equation*}
Since we know that $\varphi'(s)\geq 0$ for all $s$ and $\varepsilon>0$, we can conclude that conditions \eqref{eq:thm:relay_cond1} and \eqref{eq:thm:relay_cond2} imply that $\mathrm{div} (f_\varepsilon(x))<0$ for all $x\in \mathbb{R}^2$ and, from Bendixson-Dulac theorem \citep[Lemma 2.2]{khalil1996nonlinear}, $\dot{x}=f_\varepsilon(x)$ cannot have limit cycles. Hence, from Theorem \ref{thm:contraction_filippov} the relay feedback system from which $f_\varepsilon$ was derived cannot exhibit limit cycles.
\end{remark}
\paragraph*{Example 4}
Consider a relay feedback system \eqref{eq:relay_sys} with
\begin{equation*}
\label{eq:relayfeedback}
A=
\begin{bmatrix}
-2 & -1\\
1 & -3
\end{bmatrix},
\quad
b=
\begin{bmatrix}
1\\
3
\end{bmatrix},
\quad
c^T=
\begin{bmatrix}
0 & 1
\end{bmatrix}
\end{equation*}
Using the linear transition function $\varphi(s)=\mathrm{sat}(s)$ the corresponding regularized vector field \eqref{eq:regularized_sys} becomes
\begin{equation*}
f_\varepsilon(x)=
\begin{cases}
Ax-b & \mbox{if } c^Tx>\varepsilon\\[0.5ex]
\left(A-\dfrac{1}{\varepsilon}\,bc^T\right)x & \mbox{if } -\varepsilon<c^Tx<\varepsilon\\[0.5ex]
Ax+b & \mbox{if } c^Tx<-\varepsilon
\end{cases}
\end{equation*}
Outside $\mathcal{S}_\varepsilon$ the Jacobian of $f_\varepsilon$ is equal to $A$, and hence its measure does not depend on $\varepsilon$. On the other hand, since using the $\ell_1$-norm we have that $\mu_1(A)= \max\{-2+|1|;\;-3+|-1| \}=-1$, and $\mu_1(-bc^T)= \max\{0;\; -3+|-1| \}=0$, then when $x\in\mathcal{S}_\varepsilon$ 
\begin{equation*}
\mu\left(\frac{\partial f_\varepsilon}{\partial x}\right)  \leq \mu(A)+\frac{1}{\varepsilon}\,\mu(-bc^T)=-1.
\end{equation*}
Therefore the regularized vector field $f_\varepsilon$ remains contracting in the $\ell_1$-norm for any value of $\varepsilon$, as should  be expected since conditions of Proposition \ref{thm:contracting_relay} are satisfied in this norm. 
Hence, from Theorem \ref{thm:contraction_filippov} we can conclude that the relay feedback system taken into example is incrementally exponentially stable in the $\ell_1$-norm. In Figure \ref{fig:ex_pwa}c, we show numerical simulations of the evolution of the difference between two trajectories for this system. The dashed line is the estimated exponential decay from \eqref{eq:thm:ies_contraction_filippov} with $\bar{c}=1$ and $K=1$. An approach to contraction analysis of switched Filippov systems not requiring the use of regularization is currently under investigation and will be presented elsewhere.
\section{Conclusions}
\label{sec:conclusions}
We presented a methodology to study incremental stability in generic $n$-dimensional switched (bimodal) Filippov systems characterized by the possible presence of sliding mode solutions. The key idea is to obtain conditions for incremental stability of these systems by studying contraction of their regularized counterparts. We showed that the regularized vector field is contracting if a set of hypotheses on its modes are satisfied. In contrast to previous results, our strategy does not require explicit computation of the sliding vector field using Filippov's convex method or Utkin's equivalent control approach. Moreover, different metrics rather than the Euclidean norms can be effectively used to prove convergence. The theoretical results were applied on a set of representative examples including piecewise smooth systems, piecewise affine systems and relay feedback systems. In all cases, it was shown that the conditions we derived are simple to apply and have a clear geometric interpretation. We wish to emphasize that the tools we developed could be instrumental not only to carry out convergence analysis of Filippov systems but also to synthesize switched control actions based on their application \citep{di2015switching}.
%
%
\begin{ack}                               
SJH wishes to acknowledge support from the Network of Excellence MASTRI Materiali e Strutture Intelligenti (POR Campania FSE 2007/2013) for funding his visits to the Department of Electrical Engineering and Information Technology of the University of Naples Federico II. DF acknowledges support from the University of Naples Federico II for supporting his visits at the Department of Engineering Mathematics of the University of Bristol, U.K. The authors would like to thank the anonymous reviewers for their comments that led to a significant improvement of the manuscript.
\end{ack}

\bibliographystyle{elsarticle-harv}        
\bibliography{refs}           



\end{document}